\documentclass[a4paper,11pt]{article}
\pdfoutput=1
\usepackage{jheppub}
\usepackage[utf8]{inputenc}
\usepackage{amsmath,amssymb, latexsym}
\usepackage{xcolor}
\usepackage{graphicx}
\usepackage{hyperref}
\usepackage{mathtools}

\newcommand{\be}{\begin{equation}}
\newcommand{\ee}{\end{equation}}
\newcommand{\bea}{\begin{eqnarray}}
\newcommand{\eea}{\end{eqnarray}}

\begin{document}

\title{The Inflaton Portal to Dark Matter}

\author{Lucien Heurtier}
\emailAdd{lucien.heurtier@ulb.ac.be}
\affiliation{Service de Physique Th\'eorique - Universit\'e Libre de Bruxelles, Boulevard du Triomphe, CP225, 1050 Brussels, Belgium}
\abstract{
We consider the possibility that the inflaton is part of the dark sector and interacts with the standard model through a portal interaction with a heavy complex scalar field in equilibrium with the standard model at high energies. The inflaton and dark matter are encapsulated in a single complex field and both scalar sectors are charged under different (approximate) global U(1)'s such that the dark matter, as well as the visible pseudo-scalar are taken to be relatively light, as pseudo Nambu-Goldstone bosons of the theory. The dark matter relic density is populated by Freeze-In productions through the inflaton portal. In particular, after the reheating, production of dark matter by inflaton decay is naturally suppressed thanks to Planck stringent constraints on the dark quartic coupling, therefore preserving the non thermal scenario from any initial condition tuning. 
}
\date{\today}

{\hfill ULB-TH/17-15}
\maketitle

\vfill
\pagebreak

\section{Introduction}
The cosmological Big Bang description of our Universe suffers from (at least) two major problems. The first one concerns our misunderstanding of gravitation on galactic and extra galactic scales. Indeed, the observation of galaxies rotation curves and cluster collisions at present time remains unexplained by general relativity \cite{ZwickyRubin, Bullet, newbullet}. The second one is the puzzling quasi-homogeneity of the Cosmological Microwave Background (CMB) suggesting an unexplained fine-tuning of initial conditions of the metric fluctuations and curvature at the Big Bang time. The first problem has been addressed by invoking the existence of dark matter (DM) \cite{ZwickyRubin} while the second problem is usually assumed to be circumvented by the existence of an inflationary phase of cosmology driven by another new (scalar) particle, the inflaton \cite{Inflation1,Inflation2,Inflation3,Inflation4,Inflation5,Inflation6,Inflation7,Inflation8}. If both particles are usually considered to play a role at radically different time scales, they may nevertheless interact when the inflaton decays to produce our thermal bath.

The  usual description of dark matter dynamics can be generically divided in thermal and non-thermal scenarios \cite{portal}. The latters assume that after reheating, dark matter and the standard model have sufficient interactions to enter into thermal equilibrium before dark matter particles decouple from the thermal bath and constitute a cosmological relic. The second class of scenarios considers on the contrary that such an equilibrium never occurred and that dark matter is non-thermally produced through annihilation or decay of the visible bath, out of equilibrium. On the one hand, in both cases the interaction required to generate a quantity of dark matter at present time in agreement with astrophysical measurement should be reasonably small and usually leads to invoke the existence of an extra {\em mediator}, whose role is to encode such low interactions between the dark and the visible sectors (see e.g. \cite{Zprime,Sterile,Higgs,Zportal,Pseudoscalar}).   On the other hand, if thermal scenarios can ignore the physics of the reheating, non thermal scenario usually assume that no dark matter has been produced by inflaton decay at the time of the reheating, therefore secluding the inflationary sector from the dark sector. 

Surprisingly, the study of explicit interactions of the inflaton with low energy phenomenology has been rather poorly studied in the literature or to some extent fine-tuned through an extensive use of non minimal coupling to gravity or gauge corrections \cite{Khoze:2013uia, Choubey:2017hsq, Domcke:2017fix, Addazi:2016bus, Ferreira:2016uao, Ballesteros:2016euj, Kannike:2016jfs, Aravind:2015xst, Kahlhoefer:2015jma, Tenkanen:2016idg, Tenkanen:2016twd, Ferreira:2017ynu}.  Attempts to incorporate inflation within a context of low scale supersymmetry breaking \cite{alberto} has however shown that a complete description of low energy phenomenology of an inflation model can lead to rich constraints relating primordial cosmology to LHC physics. Furthermore interactions of the inflaton with the Brout-Englert-Higgs boson has been shown to be able to solve the question of the stability of the electroweak vacuum \cite{oleg}. However the question of the role played by the inflaton as a mediator in the dark matter production mechanism has never been explored so far. In this paper we would like to question the possibility that the inflationary sector provides a natural portal between the dark and the visible sectors without such non minimal coupling. We will for this follow as an illustrative example the steps of \cite{Boucenna:2014uma, King:2017nbl} in which dark matter is encapsulated together with the inflaton in a single complex field and interacts with the standard model through a portal interaction and show that constraints coming from both aspects of cosmology can be intimately related.

In section \ref{sec:comments} we will start by making general comments about inflation and reheating, estimating the amount of dark matter which could be produced by decay of the inflaton and sketching the difficulty of building an inflaton portal to annihilate or produce efficiently dark matter at late time.
In Section \ref{sec:Themodel} we will present the model we consider, deriving the physical spectrum and interactions of the theory. We will describe in details the inflationary potential and show that it closely constrains the coupling of the inflaton to dark matter. We will then study the non thermal production of dark matter through the inflaton portal and show that the model provides naturally a regime in which dark matter is dominantly produced by the Freeze-In mechanism rather than by direct decay of the inflaton. We will in addition compute the number of e-folds necessary to fit such a scenario with observations. Finally we will conclude and make some comments in Sec. \ref{sec:conclusion} about possible implications of the model.

\section{General comments}\label{sec:comments}

In order to dilute the metric fluctuations as well as the curvature of our Universe in sufficient proportions to match with present observations of the CMB without any initial conditions fine-tuning, the idea of single field inflation was proposed \cite{Inflation2, Inflation3, Inflation5, Inflation6}.  The inflationary action can simply be written
\be\label{action_inflation}
S=\int d^4 x \sqrt{-g}\left[\frac{1}{2}R+\frac{1}{2}g^{\mu\nu}\partial_{\mu}\phi\partial_{\nu}\phi-V(\phi)\right]\,,
\ee
In this paradigm, the classical rolling of a scalar field $\phi$ -- called the inflaton -- along its potential $V(\phi)$ triggers the acceleration of the Universe expansion until it  falls down into the vacuum. While the field rolls towards its minimum $\phi_0$, the value of the potential $V(\phi)$ dominates the energy density of the Universe and behaves as a friction term in the Friedmann-Lema\^itre equations, such that the rolling is slowed down. This {\em slow-roll} regime is crucial to suppress kinetic energy as compared to potential energy -- rendering the effect of inflation similar to the one of a cosmological constant -- and to ensure that inflation lasts long enough before the scalar field starts oscillating around the vacuum. This regime is ensured as long as
\bea
\epsilon_V&\equiv& \frac{M_p^2}{2}\left(\frac{V_{,\,\phi}}{V}\right)^2 < 1\label{SR1}\,,\\
|\eta_V|&\equiv& \left|M_p^2\frac{V_{,\,\phi\phi}}{V}\right|<1\,, \label{SR2}
\eea
where $\epsilon_V$ and $\eta_V$ are called the slow roll parameters. These parameters are the key ingredients, being given an inflationary model, to compute the observables measured by the Planck collaboration, that are the tensor-to-scalar ratio $r$ and the spectral index $n_s$. These are computed at the time of horizon crossing (denoted by a star), usually taken to happen 50 and 60 e-folds before the slow roll regime ends
\bea
n_s&\approx&1 -6\ \epsilon_{V}^{\star}+2\ \eta_{V}^{\star}\nonumber\\
r&\approx&16\ \epsilon_{V}^{\star}\,.
\eea
Furthermore, the energy scale of inflation is strongly constrained by the normalization of the scalar perturbations power spectrum amplitude \cite{Ade:2015xua}
\begin{equation}
A_s \equiv \frac{V(\phi^{\star})}{24\pi^2\ \epsilon_V^{\star} M_p^4}\approx 2.198\times 10^{-9}\,,
\end{equation}
where $M_p$ denotes the reduced Planck mass.
In the simplest models of inflation, such as the {\em chaotic} scenario $V(\phi)= m^2\phi^2/2$, such condition fixes the mass of the inflaton to $m\sim 10^{13}$ GeV and asking for 50-60 efolds of inflation requires the inflaton to roll over transplanckian values $\Delta \phi \sim \mathcal O (10 M_p)$. These are generic features of a wide class of inflationary models called {\em large field inflation} models. Although a plethora of other inflation models exist in the literature we will focus in the following on this class of scenarios.

After the inflation era ends (once the slow roll parameters reach unity) the friction term becomes subdominant and the field $\phi$ starts rolling down his potential classically thus oscillating around the vacuum. Once the Hubble scale is of order the decay width of the inflaton
\begin{equation}
H\sim \Gamma_{\phi}\,,
\end{equation}
the decay of the inflaton becomes efficient enough to {\em reheat} the Universe, by populating a thermal bath of relativistic particles with $g_{\star}$ degrees of freedoms. The reheating temperature is thus defined as
\be\label{eq:defreheating}
	\rho_{\phi}=3H^2M_p^2=3\Gamma_\phi^2M_p^2=\frac{\pi^2 g_*}{30}T_R^4\,.
\ee
For $g_{\star}$ of order a hundred this provides the estimation $T_R\sim 0.5\sqrt{\Gamma_\phi M_p}$. Therefore, having a complete and explicit description of the interactions between the inflaton, dark matter and the standard model fixes the reheating temperature as a function of the model parameters.

We can at this point make a few important remarks concerning the production of dark matter. On the one hand the very high mass of the inflaton favoured by large field inflation models will inevitably suppress any cross section of annihilation through exchange of an inflaton particle. In order to produce thermally (or non thermally) dark matter, one is thus constrained to use large values of the couplings not to overclose the universe (or to produce enough dark matter). On the other hand, pushing up the couplings of the inflaton to the visible or dark sector will automatically increase the reheating temperature. Even if no generic bound is known about the reheating temperature, a too low or too high reheating temperature can have implications for the upcoming cosmological history of the Universe. A thermal description of leptogenesis \cite{Fukugita:1986hr} can for instance only be realized for large reheating temperatures $T_R\gtrsim 10^9$ \cite{Buchmuller:2004nz} GeV whereas in a supersymmetric framework the thermal production of gravitinos or the emergence of long lived particles destroying the Big Bang nucleosynthesis lead to various upper bounds on the reheating temperatures \cite{Fayet:1981sq, Pagels:1981ke, Weinberg:1982zq, Ellis:1984er, Kawasaki:2004yh, Kawasaki:2004qu, Jedamzik:2006xz, Moroi:1993mb, Pradler:2006hh, Asaka:2000zh, Steffen:2008bt, Covi:2010au, Roszkowski:2012nq}. Asking for a reheating temperature lower than the inflation scale, we will focus in the following on non-thermal dark matter production, since it requires lower interactions of DM with the standard model.

Finally, opening the possibility that the inflaton interacts with both sectors permits that the inflaton decays directly into dark matter during the reheating. Assuming that this happens instantaneously, and that the inflaton couples to pairs of dark matter particles with a branching ratio \begin{equation}Br(\phi\to DM,DM)=\frac{\Gamma(\phi\to DM,DM)}{\Gamma_{\phi}^{tot}}\,,\end{equation}one can estimate the amount of dark matter produced immediately after the reheating by
\begin{equation}\label{eq:analyticReheating}
\frac{\rho_{DM}}{m_{DM}}(T=T_R)=2Br(\phi\to DM,DM)\frac{\rho_{\phi}}{m_{\phi}}(T=T_R)\,.
\end{equation}
Using entropy conservation, one can derive the relic density corresponding to this production during the reheating to be
\begin{eqnarray}\label{eq:DecayRelic}
&&\Omega_{DM}h^2\approx10^{-4}\frac{m_{DM}}{m_{\phi}}Br(\phi\to DM,DM)\left(\frac{T_R}{T_0}\right)\,,\nonumber\\
&\approx&0.1\times\left(\frac{10^{13}\mathrm{GeV}}{m_{\phi}}\right)\left(\frac{m_{DM}}{\mathrm{MeV}}\right)\left(\frac{\mathrm{eV}}{T_0}\right)\left(\frac{T_R}{10^{10}\mathrm{GeV}}\right)\nonumber\\
&\times&Br(\phi\to DM,DM)\,.
\end{eqnarray}
Thus it is straightforward to note that, even for a rather small mass of dark matter and a reasonable value of the reheating temperature, a direct decay of the inflaton into dark matter may overclose the universe in certain regimes. While this would produce dark matter which would be far too warm to explain structure formation, such a production at the reheating time would render inappropriate a non thermal description of dark matter production. We will thus in what follows systematically take care that the decay production during reheating is sub-dominant as compared to the non thermal process. We will hence require that the branching ratio into dark matter is sufficiently small to suppress the reheating contribution to the relic density at low energy
\begin{equation}
Br(\phi\to DM,DM)\ll 1\,.
\end{equation}

\section{The inflaton portal model}\label{sec:Themodel}

In this section we propose a model in which the visible and invisible sectors are constituted of two complex scalar fields charged under different global $U(1)$'s. The breaking of these $U(1)$'s -- at scales which will be made explicit in the following -- will give birth to two Nambu-Goldstone bosons both in the visible and the dark sectors whose masses will be protected by the latter approximate symmetries. Finally the inflaton -- the scalar field in the dark sector -- will interact with dark matter (the pseudo Nambu Goldstone boson of the hidden sector) through its quartic coupling and vacuum expectation value (vev) $\lambda_{\phi} v_{\phi}$ whose value will be constrained by inflationary observables to be highly suppressed. Such suppression will guarantee that dark matter is safely produced by a non thermal Freeze-In and not by the reheating itself.

The lagrangian we consider is\footnote{Note that the choice of the last term where the vev's cancel after spontaneous breaking is equivalent, after a field redefinition, to a term $\delta (\Phi^{\dagger}\Phi) (\Sigma^{\dagger}\Sigma)$, as long as the bare masses are related to the vev's as in \eqref{eq:vevs}.}
\bea \label{eq:lagrangian}
-\mathcal L&\supset& \frac{\lambda_{\phi}}{2}\left(\Phi^{\dagger}\Phi-\frac{v_{\phi}^2}{2}\right)^2+\frac{\lambda_{\sigma}}{2}\left(\Sigma^{\dagger}\Sigma-\frac{v_{\sigma}^2}{2}\right)^2\nonumber\\
&+&{\delta}\left(\Phi^{\dagger}\Phi-\frac{v_{\phi}^2}{2}\right)\left(\Sigma^{\dagger}\Sigma-\frac{v_{\sigma}^2}{2}\right)\,,
\eea
in which one can expand fields around there minimum of potential
\begin{eqnarray}
\Phi&=& \frac{1}{\sqrt 2}(v_{\phi}+\phi +i S_d)\,,\nonumber\\
\Sigma&=& \frac{1}{\sqrt 2}(v_{\sigma}+\sigma +i S_v)\,.
\end{eqnarray}

Under these notations, the scalar $\phi$ will be playing the role of the inflaton whereas the pseudo scalar $S_d$ is the dark matter constituent and $(\sigma, S_v)$ are assumed to be in thermal equilibrium with the standard model after reheating.

The masses of $\phi$ and $\sigma$ are related to the vacuum expectation values $v_{\phi}$ and $v_{\sigma}$ according to the equations of motion
\begin{equation}
\label{eq:vevs}
m_{\phi}^2=\lambda_{\phi}v_{\phi}^2 - \delta v_{\sigma}^2\,,\quad\text{and}\quad m_{\sigma}^2=\lambda_{\sigma}v_{\sigma}^2 - \delta v_{\phi}^2\,.
\end{equation}

\begin{figure}
\includegraphics[width=\linewidth]{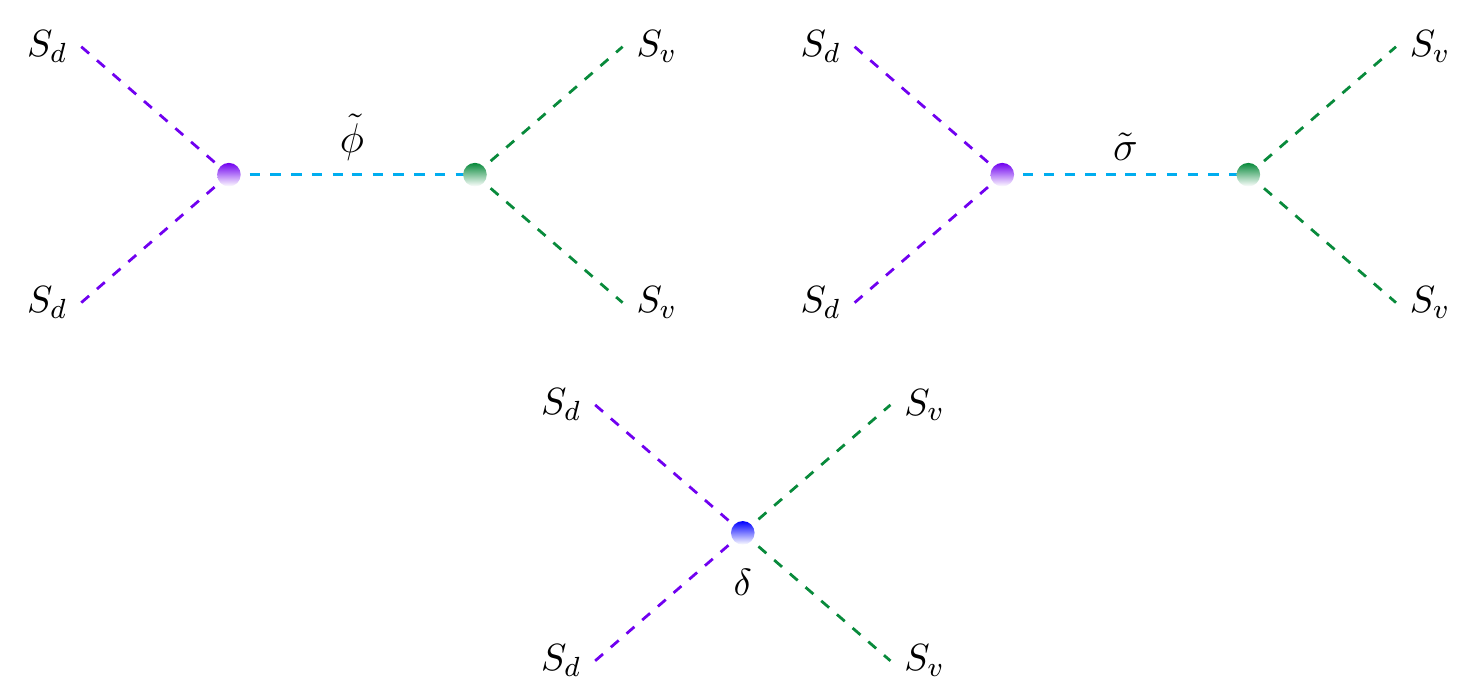}
\caption{\label{fig:UVchannels} Diagrams after rotation into the physical basis $(\tilde \phi,\tilde \sigma)$ leading to annihilation of dark matter into the visible sector.}
\end{figure}

Expanding the lagrangian around the vacuum, one can write
\begin{eqnarray}\label{eq:lagrangianexpand}
-\mathcal L&\supset& \frac{\lambda_{\phi}}{8}\left(\phi^4+4v_{\phi}^2\phi^2+4v_{\phi}\phi^3+S_d^4+4v_{\phi}\phi S_d^2+2\phi^2S_d^2\right)\nonumber\\
&+& \frac{\lambda_{\sigma}}{8}\left(\sigma^4+4v_{\sigma}^2\sigma^2+4v_{\sigma}\sigma^3+S_v^4+4v_{\sigma}\sigma S_v^2+2\sigma^2S_v^2\right)\nonumber\\
&+&\frac{\delta}{4}\left((\phi+v_{\phi})^2(\sigma+v_{\sigma})^2+S_v^2S_d^2+S_v^2(\phi+v_{\phi})^2\right.\nonumber\\
&&\quad+\left.S_d^2(\sigma+v_{\sigma})^2\right)\,,
\end{eqnarray}
in which the inflaton portal interaction $\delta (\Phi^{\dagger}\Phi-\frac{v_{\phi}^2}{2})(\Sigma^{\dagger}\Sigma-\frac{v_{\sigma}^2}{2})$ generates three possible channels of annihilation $S_d,S_d \rightarrow S_v, S_v$ as depicted in Fig. \ref{fig:UVchannels}.
Diagonalizing the mass matrix
\begin{equation}
\frac{1}{2}\mathcal M^2=\frac{1}{2}\begin{pmatrix}
m_{\phi}^2&m_{\phi\sigma}^2\\
m_{\phi\sigma}^2&m_{\sigma}^2
\end{pmatrix}\,,
\end{equation}
where we defined $m_{\phi\sigma}^2=\delta v_{\phi}v_{\sigma}$, one obtains the masses
\begin{eqnarray}\label{eq:masses}
m_{\tilde\phi,\tilde\sigma}^2=\frac{m_{\phi}^2+m_{\sigma}^2}{2}\mp\left(\left(\frac{m_{\phi}^2-m_{\sigma}^2}{2}\right)^2+m_{\phi\sigma}^4\right)^{1/2}
\end{eqnarray}
associated to the eigenvectors
\begin{eqnarray}
\phi=c_{\theta}\ \tilde\phi-s_{\theta}\ \tilde\sigma\,,\nonumber\\
\sigma=c_{\theta}\ \tilde\sigma+s_{\theta}\  \tilde\phi\,,
\end{eqnarray}
where the diagonalizing cosines, defining the parameter $a=(m_{\phi}^2-m_{\sigma}^2)/2m_{\phi\sigma}^2$, are given by
\begin{eqnarray}
c_{\theta}&=&\frac{1}{\sqrt 2}\left(1-\frac{a}{\sqrt{1+a^2}}\right)^{1/2}\,,\nonumber\\
s_{\theta}&=&\frac{1}{\sqrt 2}\left(1+\frac{a}{\sqrt{1+a^2}}\right)^{1/2}\,.
\end{eqnarray}

In order to get positive physical masses, the mass scale $m_{\phi\sigma}$ should satisfy the condition $m_{\phi\sigma}<\sqrt{m_{\phi}m_{\chi}}$.

\subsection{Inflation}

During inflation, the field $\sigma$ is imposed to be sufficiently massive not to interfere with the single field dynamics of inflation
\begin{equation}
m_{\sigma}^2\gtrsim H^2\,.
\end{equation}
Furthermore, the pseudo-goldstone $S_d$, $S_v$, as well as the standard model are assumed to get effective Hubble scale masses during inflation, such that they remain stabilized at zero field values during inflation\footnote{Note that fields direction remaining flat for large inflaton values remain stabilized to zero due to the High Hubble friction they undergo in a similar manner than hubble scale mass fields do. Other fields interacting with the inflaton get effective masses due to the large field values of the inflaton during inflation $\phi\sim M_p$.}. In other words, very light excitations get immediately diluted by inflation and never populate the universe before fast expansion ends.

Under such assumption one can integrate out the field $\Sigma$ in the lagrangian  \eqref{eq:lagrangian} and set the goldstones to zero to obtain the inflation potential
\begin{equation}
V_{inf}(\phi)=\frac{\lambda_{\phi}-\delta^2/\lambda_{\sigma}}{8}\left(\phi^4+4v_{\phi}\phi^2+4v_{\phi}\phi^3\right)\,.
\end{equation}

\begin{figure}
\includegraphics[width=\linewidth]{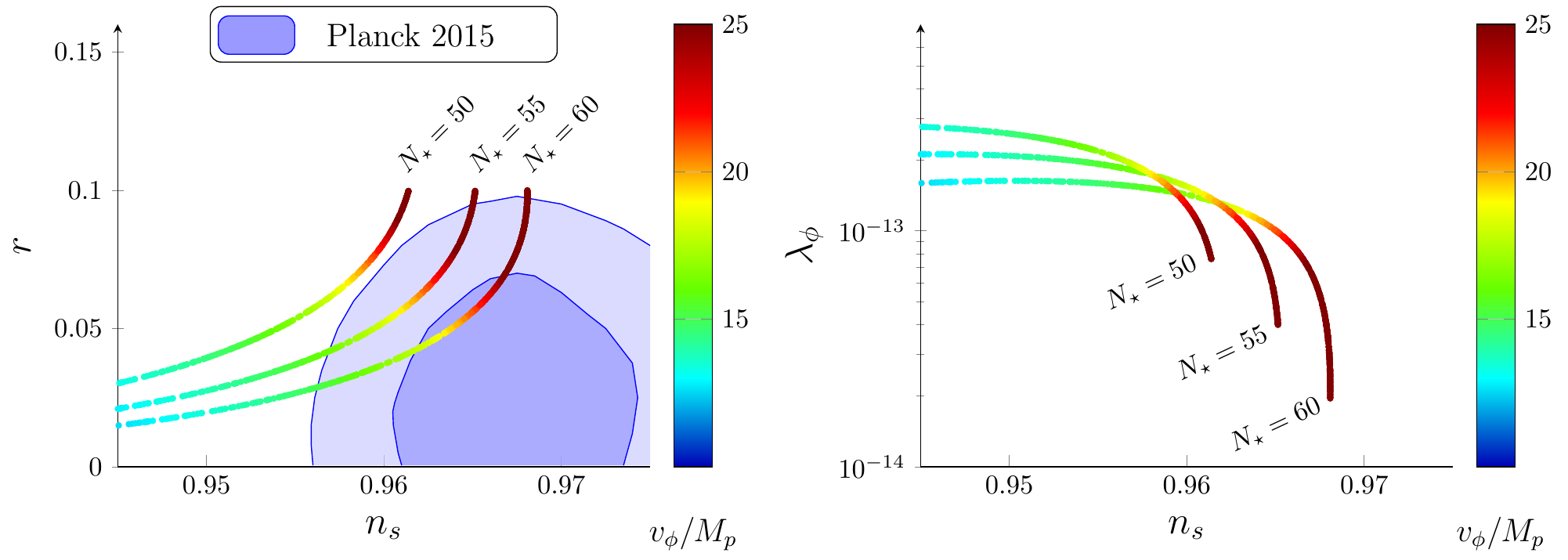}
\caption{\label{fig:inflationgeneral} Left panel : Inflation observables for various numbers of e-folds and arbitrary value of the vev $v_{\phi}$ (colorbar). Right panel : Value of the quartic coupling $\lambda_{\phi}$ imposed by the COBE normalization, as a function of the tilt $n_s$ and the inflaton vev $v_{\phi}$ (colorbar).}
\end{figure}

This potential is similar to those of so called {\em new inflation scenario} studied in \cite{Kallosh:2010ug, Kallosh:2007wm, Kawasaki:2001as}.in which the inflaton can roll down its potential from the origin $\Phi\gtrsim 0$ to the vacuum $\Phi\approx v_{\phi}$. Such scenarios, sometime denominated by "quartic hilltop" can lead to a tensor to scalar ratio $r$ and a spectral tilt $n_s$ in agreement with Planck measurements  \cite{Ade:2015lrj}, as is depicted in Fig. \ref{fig:inflationgeneral} asking for 50 to 60 e-folds of inflation.

Generically, asking that the model prediction lay inside the 2-$\sigma$ contour of such constraints impose that $\lambda_{\phi}\sim 10^{-13}$ and $v_{\phi}\sim 20 M_p$, which we will use as reference points in the following analysis. 

As we will see in the next section, asking for the model to produce the right amount of dark matter will impose that $\delta\sim 10^{-11}$. For a coupling constant $\lambda_{\sigma}$ of order $\mathcal{O}(0.1)$ this leads to a correction of the quartic coupling of order $10^{-21}\lesssim \lambda_{\phi}$.

Finally these numbers will be refined in the last section after a proper study of the reheating temperature and the number of e-folds necessary to release inflation.

\subsection{Dark matter production}

As announced, our model will provide a natural way to produce dark matter non-thermally, thanks to the inflaton portal interaction. 

In our set up, the two scalars $\phi$ and $\sigma$ have masses well above the reheating temperature $m_{\sigma}\gg m_{\phi}\gtrsim T_R$ such that their dynamics doesn't influence the cosmological history once reheating occurs. The two pseudo-scalars $S_v$ and $S_d$, as Nambu Goldstone bosons of the theory, have masses protected by the (approximate) global U(1) symmetries under which they are charged. As they acquire tiny masses through gravitational effects, we will consider their masses $m_v$ and $m_d$ as -- arbitrarily small -- free parameters of the model.

Considering Eq.~\eqref{eq:vevs}, one can immediately make useful remarks about the consistency of the model construction. On the one hand, asking that the scalar $\sigma$ can be integrated out during inflation imposes that
\begin{equation}\label{eq:approx1}
m_{\sigma}\gtrsim H\sim 10^{14}\mathrm{GeV}\,.
\end{equation}
Assuming the coupling $\lambda_{\sigma}$ to be close to unity thus imposes that $v_{\sigma}\gtrsim 10^{14}\mathrm{GeV}$. On the other hand, since the inflaton mass is fixed by the normalization condition to be of order $m_{\phi}\sim 10^{13}$ GeV, we will be able to work in the limit where
\begin{equation}\label{eq:approx2}
\delta \ll 10^{-9}\lesssim \lambda_{\phi}\frac{m_{\sigma}^2}{m_{\phi}^2}\,.
\end{equation}
Under such condition, the physical masses of the fields are related to their vev by the simple relations
\begin{equation}
m_{\tilde \phi}^2\approx m_{\phi}^2\approx\lambda_{\phi}v_{\phi}^2\,,\quad\text{and}\quad m_{\tilde \sigma}^2\approx m_{\sigma}^2\approx\lambda_{\sigma}v_{\sigma}^2\,.
\end{equation}
With such low mixing $\delta$ one obtains $m_{\sigma}^2\gg m_{\phi}^2,m_{\phi\sigma}^2$. This gives $|a|\approx m_{\sigma}^2/2m_{\phi\sigma}^2\gg 1$ and
\begin{equation}\label{eq:cos}
c_{\theta}\approx 1\,,\quad\text{and}\quad s_{\theta}\approx \frac{1}{2a}\ll 1\,.
\end{equation}
After rotation into physical states in the lagrangian \eqref{eq:lagrangianexpand}, the couplings of the scalars to the visible and dark sectors are given by
\begin{eqnarray}\label{eq:couplings}
\tilde\phi S_d^2 &:& \frac{\lambda_{\phi}v_{\phi} c_{\theta}+\delta v_{\sigma} s_{\theta}}{2}\approx \frac{\lambda_{\phi}v_{\phi}}{2}\nonumber\\
\tilde\phi S_v^2 &:& \frac{\lambda_{\sigma}v_{\sigma} s_{\theta}+\delta v_{\phi} c_{\theta}}{2}\approx \delta v_{\phi}\nonumber\\
\tilde\sigma S_v^2 &:& \frac{\lambda_{\sigma}v_{\sigma} c_{\theta}-\delta v_{\phi} s_{\theta}}{2}\approx \frac{\lambda_{\sigma}v_{\sigma}}{2}\nonumber\\
\tilde\sigma S_d^2 &:& \frac{-\lambda_{\phi}v_{\phi} s_{\theta}+\delta v_{\sigma} c_{\theta}}{2}\approx \frac{\delta v_{\sigma}}{2}\,,
\end{eqnarray}
where the right side approximations are making use of Eq. \eqref{eq:approx1} and \eqref{eq:approx2}.

The mediation between the dark and the visible sector is thus assured by the exchange of the inflaton $\phi$, as well as the scalar $\sigma$, to which is added a contact term $(\delta/4) S_d^2 S_v^2$. Therefore, the inflaton portal term generates three different diagrams whose squared amplitude is
\begin{eqnarray}
|\mathcal{M}|^2&\approx&\left(\frac{2\times 2}{4}\right)^2\left[\ \frac{2\delta \lambda_{\phi}v_{\phi}^2}{m_{\tilde\phi}^2}+\ \frac{\delta \lambda_{\sigma}v_{\sigma}^2}{m_{\tilde\sigma}^2}+\delta\ \right]^2\approx 16\ \delta^2\,.\nonumber\\
\end{eqnarray}

Assuming that the pseudo scalar $S_v$ is in thermal equilibrium with the visible bath and produces dark matter out of equilibirum (at energies lower than $m_{\phi}$), one can compute the Freeze-In relic density using 
\begin{eqnarray} \label{eq:NT1}
\langle\sigma v\rangle n_{eq}^2 &=&\frac{4\pi}{2}\frac{T}{32(2\pi)^6}\int_{4m_d^2}^{\infty}|\mathcal M|^2\sqrt{s-4m_d^2}K_1(\sqrt s/T)ds\,,\nonumber\\
\end{eqnarray}
and the approximate Boltzmann equation for non thermal production\footnote{Note that the last equality holds only because the squared amplitude is a constant in our case.} \cite{portal, Chu:2013jja, Mambrini:2013iaa, Mambrini:2015vna}
\begin{equation}\label{eq:NT2}
\frac{d Y}{d x}=\frac{1}{x  H\ {\bf s}}\langle\sigma v\rangle n_{eq}^2=\left(\frac{45}{\pi}\right)^{3/2}\frac{|\mathcal M|^2M_p}{(2\pi)^7 g_s\sqrt{g_{\rho}} 8}\frac{x^2}{m_d}K_1(x)^2\,.
\end{equation}

We thus get for the relic density generated by non thermal production
\begin{equation}
\Omega_{non-therm.}h^2\approx 0.12\times\left(\frac{\delta}{5.1\times 10^{-12}}\right)^2\,.
\end{equation}
Using this value, together with $\lambda_{\sigma}\sim 0.1$ and $v_{\phi}\sim v_{GUT}\sim 10^{16} $ GeV, one can check that the hypothesis \eqref{eq:approx1} and \eqref{eq:approx2} are verified a posteriori.

\subsection{Reheating}

The minimality of the set up and explicit coupling of the inflaton to the visible and invisible sector provides an unambiguous way to derive its partial and total decay width of decay into dark matter and standard model particles.

Using the couplings \eqref{eq:couplings} one gets
\begin{eqnarray}
\Gamma_{\tilde \phi\to S_d,S_d}&=&\frac{1}{32\pi m_{\tilde\phi}^2}\left(\frac{\lambda_{\phi}v_{\phi} c_{\theta}+\delta v_{\sigma} s_{\theta}}{2}\right)^2\sqrt{m_{\tilde\phi}^2-4m_d^2}\,,\nonumber\\
&\approx& \frac{1}{32\pi m_{\phi}}\left(\frac{\lambda_{\phi}v_{\phi}}{2}\right)^2\nonumber\\
\Gamma_{\tilde \phi\to S_v,S_v}&=&\frac{1}{32\pi m_{\tilde\phi}^2}\left(\frac{\lambda_{\sigma}v_{\sigma} s_{\theta}+\delta v_{\phi} c_{\theta}}{2}\right)^2\sqrt{m_{\tilde\phi}^2-4m_d^2}\,,\nonumber\\
&\approx& \frac{1}{32\pi m_{\phi}}\left(\delta v_{\phi}\right)^2\,.\nonumber\\
\end{eqnarray}
Thus the branching ratio of decay into dark matter
\begin{equation}
Br(\phi\to S_d,S_d)=\frac{\lambda_{\phi}^2}{\lambda_{\phi}^2+4\delta^2}\approx \frac{\lambda_{\phi}^2}{4\delta^2}\sim 4\times 10^{-4}\,,
\end{equation}
 is naturally suppressed, due to the constraint imposed on the one hand by inflationary measurement on $\lambda_{\phi}~\sim~10^{-13}$, and on the other hand by the relic density experimental value $\delta \sim 5\times10^{-12}$.
 
\begin{figure}
\includegraphics[width=\linewidth]{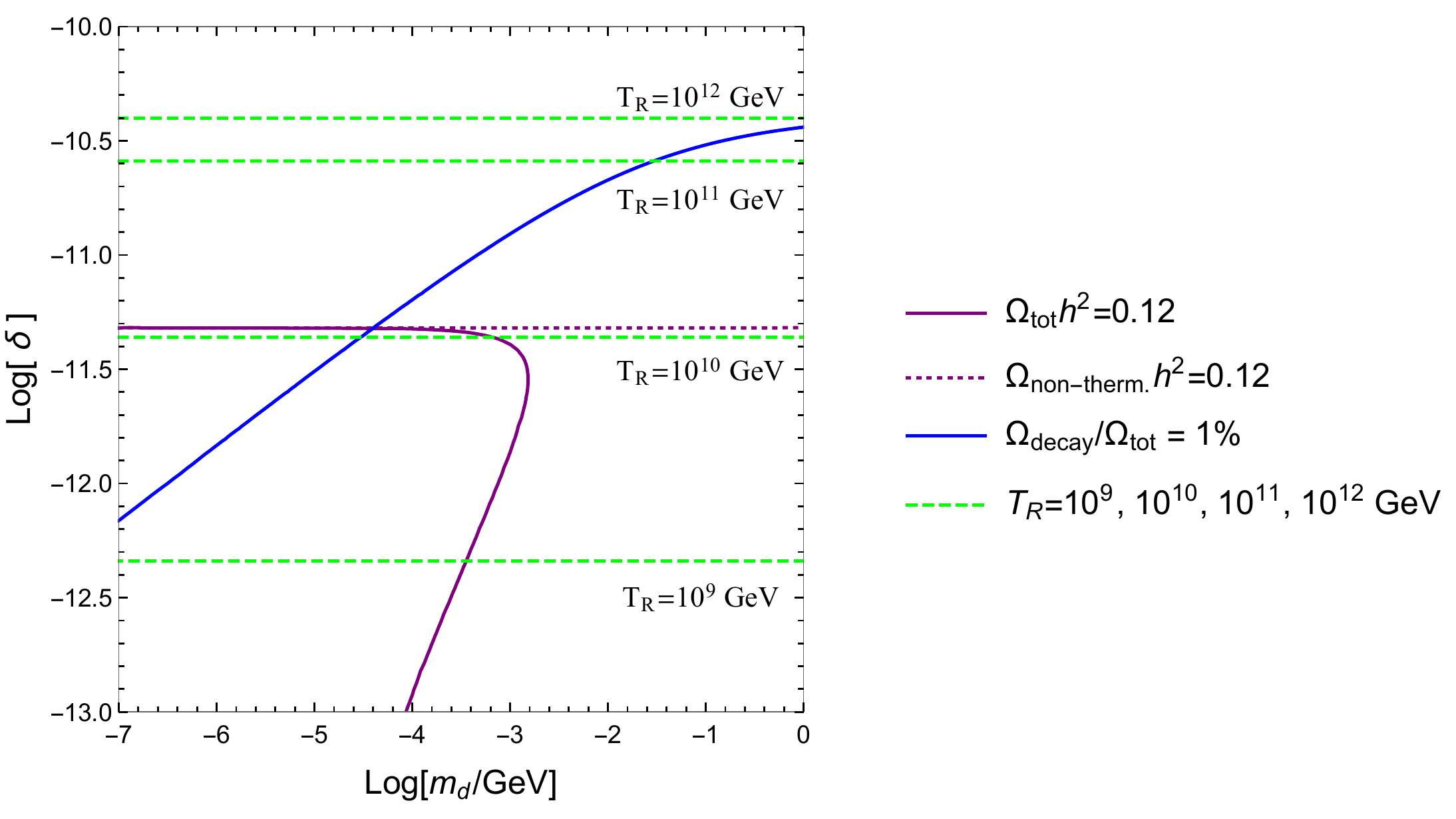}
\caption{\label{fig:finalplot} Inflaton portal mixing $\delta$ as a function of the dark matter mass $m_d$ such that the scalar $S_d$ constitute the only dark matter constituent of the Universe, where we fixed $v_{\sigma}=10^{16}$ Vev, $\lambda_{\sigma}=0.1$ and $(\lambda_{\phi},v_{\phi})=(10^{-13},20M_p)$ to match with experimental measurement of the inflationary observables. The purple full line represent the total relic density whereas the purple dotted line stands for only the non thermal contribution to the latter. The blue line delimits the region where the non thermal contribution constitute more than 99\% of the total relic density (upper part of the plot). Finally the green dashed line indicate the values of the reheating temperature $T_R$ for different values of the parameters.}
\end{figure}

However, such a suppression may not be sufficient in the case where the dark matter would be massive enough. Indeed, the relic density produced by a direct decay of the inflaton is proportional to the dark matter mass, according to Eq.~\eqref{eq:DecayRelic}. As is depicted in Fig. \ref{fig:finalplot}, requiring that the direct production of dark matter through decay of the inflaton does not represent more than 1\% of the total relic density thus leads to an upper bound on the dark matter mass
\begin{equation}
\frac{\Omega_{decay}}{\Omega_{decay}+\Omega_{non-therm.}}<1\%\ \longrightarrow\ m_d \lesssim 40\ \mathrm{keV}\,.
\end{equation}

The reheating temperature roughly corresponds to the temperature of the relativistic thermal bath when produced at the time where the decay of the inflaton starts competing with expansion $H\sim \Gamma_{\phi}$\cite{Kofman:1994rk}

\bea\label{eq:reheatingdef}
	\rho_{re}&=&3H_{re}^2M_p^2=3\Gamma_\phi^2M_p^2\equiv\frac{\pi^2 g_*}{30}T_R^4, \nonumber\\
	 &&\Rightarrow\ T_R\approx 0.5 \sqrt{\Gamma_{\phi}M_p}\,,
\eea
where $g_*\sim 100$ is the number of relativistic degrees of freedom at the reheating time.
As is depicted in Fig. \ref{fig:finalplot} the value predicted by our scenario is of order $T_R \sim 10^{10}\mathrm{GeV}\lesssim m_{\phi}$, thus in agreement with the assumptions used while discussing non thermal production. As indicated in introduction, if sufficiently high regarding thermal leptogenesis \cite{Buchmuller:2004nz}, note that in the context of supersymmetric set up, such value of the reheating temperature may be in tension with other existing bounds \cite{alberto, Dudas:2017rpa, Benakli:2017whb, Fayet:1981sq, Pagels:1981ke, Weinberg:1982zq, Ellis:1984er, Kawasaki:2004yh, Kawasaki:2004qu, Jedamzik:2006xz, Moroi:1993mb, Pradler:2006hh, Asaka:2000zh, Steffen:2008bt, Covi:2010au, Roszkowski:2012nq}.

\begin{figure}
\includegraphics[width=\linewidth]{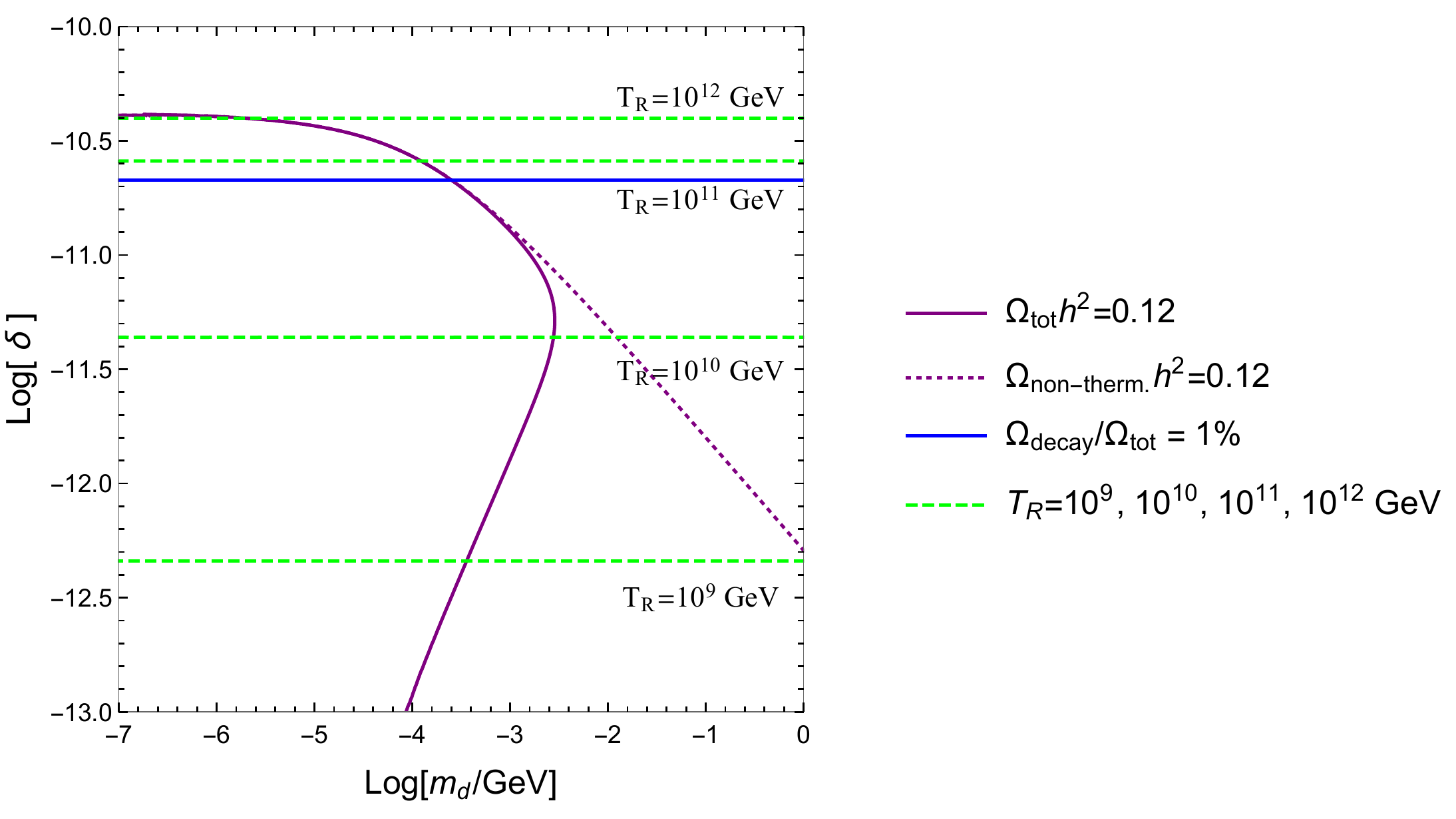}
\caption{\label{fig:finalplot2} Same legend as Fig. \ref{fig:finalplot} for a mass $m_v=10$ MeV.}
\end{figure}

Note that for such a light dark matter to be produced non thermally till $x\equiv m_d/T\approx 20$, the visible scalar $S_v$ should be in equilibrium with the thermal bath after the neutrino decoupling temperature, which would lead to strong deviation of the effective number of relativistic degrees of freedom as compared to its experimental value \cite{Boehm:2012gr}. In order to avoid such problem, one has to assume that the mass of the pseudo scalar $S_v$ is higher than $T_{\nu}\sim $ MeV. Thus the non thermal production is forced to stop before the temperature reaches the dark matter mass. This effect is taken into account in Fig. \ref{fig:finalplot2} where we fixed the mass $m_v=10$ MeV. One can see in this case that the parameter $\delta$ should be slightly higher, but still in agreement with conditions \eqref{eq:approx1} and \eqref{eq:approx2}. This increase of the inflaton portal interaction therefore leads to an increase of the possible dark matter mass allowed since the threshold is now fixed by the visible mass $m_d$, permitting the dark matter mass to be as high as $m_d \lesssim 250\ \mathrm{keV}$.

\subsection{Number of e-folds}
So far we have considered the number of e-fold as a free parameter of the model. However, the latter strongly depends on the physics of the reheating and inflation\cite{Liddle:2003as, Munoz:2014eqa}. In our set up, the reheating temperature is determined by Eq. \eqref{eq:reheatingdef} and can be computed for any choice of parameters. As we will see, it will constrain the number of e-fold and therefore the inflationary observables.

\begin{figure}
\includegraphics[width=\linewidth]{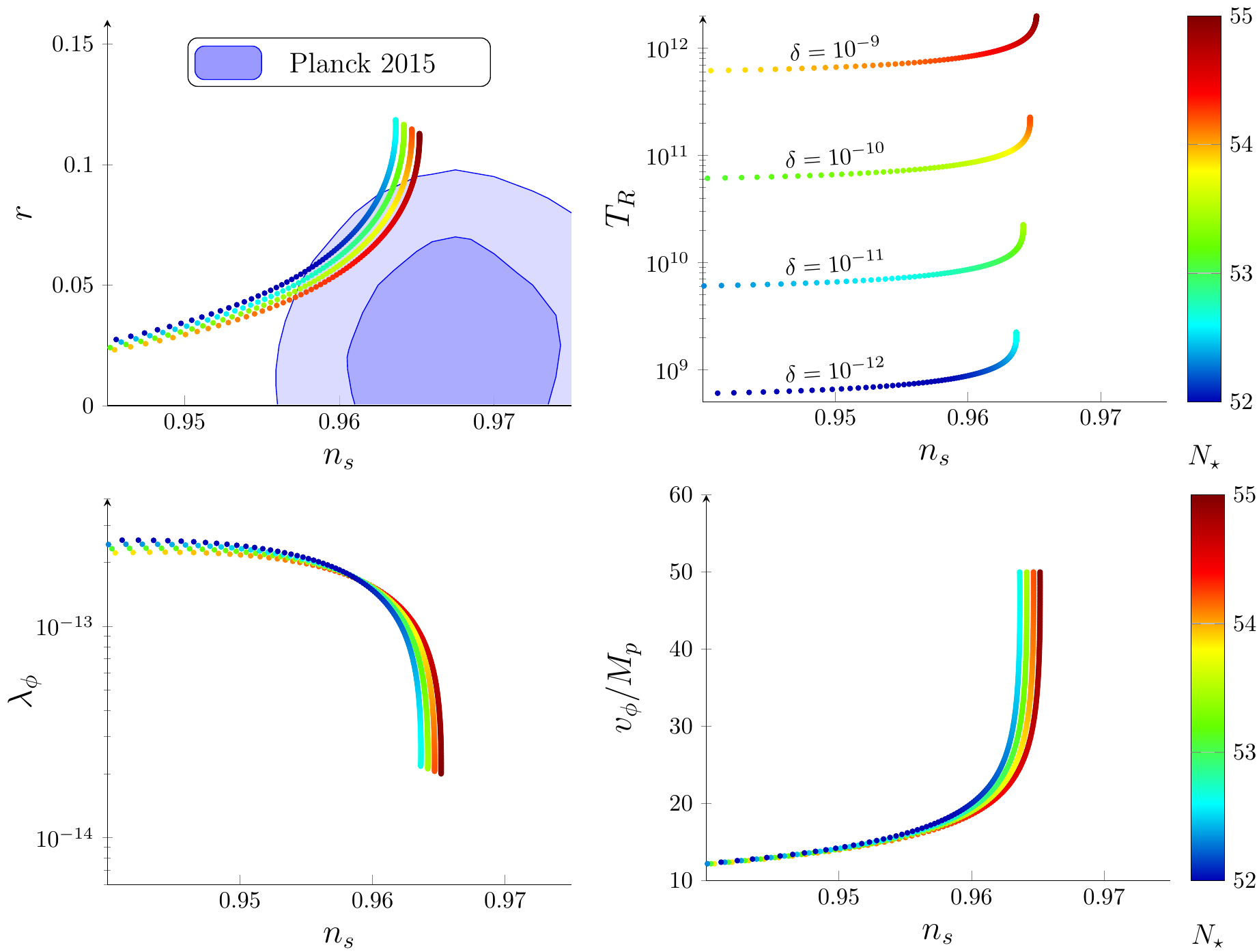}
\caption{\label{fig:reheating}From Top-left to bottom-right : Tensor-to-scalar ratio, reheating temperature, quartic coupling and inflaton vev, as functions of the spectral tilt $n_s$, and parametrized by the number of e-folds $N_{\star}$ for different choices of $\delta$.}
\end{figure}

Following \cite{Liddle:2003as} we can define the number of e-folds between the end of inflation and the reheating time $N_{re}$ by
\be
	\rho_{re}=\rho_{end}e^{-3N_{re}(1+\bar{w}_{re})}=\frac{\pi^2 g_*}{30}T_{re}^4,
\ee
where $\rho_{re}$ is the energy scale after the reheating, $g_*~\sim~100$ is the number of relativistic degrees of freedom of the thermal bath, $\bar w_{re}$ refers to the equation of state during the reheating and $\rho_{end}$ stands for the energy density at the end of inflation.  To fix ideas we will stick to the case $\bar{w}_{re}=0$, whereas a fully general study could in principle cover the range $-1/3<\bar{w}_{re}<1/3$. However, note that the idealistic case of an {\em instantaneous} reheating would sit at the crossing of this class of models and that the range $0.25\gtrsim\bar{w}_{re}\gtrsim 0$ is favoured by the literature on reheating \cite{Turner:1983he, Munoz:2014eqa}. Relating the pivot scale $k$ at which the CMB is observed  to the scale of inflation ($a_k H_k$), one can write \cite{Liddle:2003as}
\be
\label{eq_reh_as}
	\frac{k}{a_0H_0}=\frac{a_k}{a_{end}}\frac{a_{end}}{a_{re}}\frac{a_{re}}{a_{eq}}\frac{a_{eq} H_{eq}}{a_0 H_0}\frac{H_k}{H_{eq}},
\ee
where the subscripts `$end$', `$eq$' and `0' denote the end of inflation, matter-radiation equality and the current time respectively. Using this decomposition, one can relate the total number of e-folds during inflation to the few e-folds covered till the reheating time\cite{Munoz:2014eqa}
\bea 
\label{eq_reh_Nstar}
	N_{\star}&=&\frac{1}{4} N_{re}-\log{\frac{k}{a_0 T_0}}-\frac{1}{4}\log{\frac{30}{g_*\pi^2}}-\frac{1}{3}\log{\frac{11 g_*}{43}}\nonumber\\
	&-&\frac{1}{4}\log{\frac{3}{2}\frac{V_{end}}{M_p^4}}+\frac{1}{2}\log{\frac{\pi^2 r A_s}{2}}.
\eea
This implicit equation where $r$ depends on $N_{\star}$
can be solved numerically to obtain the number of e-fold $N_{\star}$ as a function of the reheating temperature. The latter depends mainly on three parameters : the inflaton vev and quartic couplings $(v_{\phi},\lambda_{\phi})$ and the inflaton portal coupling $\delta$. As a matter of fact, the normalization condition relates the coupling $\lambda_{\phi}$ to the vev of the inflaton, and we have seen that the relic density constraint fixes the value of the coupling $\delta$, depending on the dark matter and visible masses. For a given choice of $\delta$, the relation \eqref{eq_reh_Nstar} together with the normalization condition will thus relate the three parameters $(N_{\star},v_{\phi},\lambda_{\phi})$ and provide one to one relations that we can parametrize by $N_{\star}$. In Fig. \ref{fig:reheating} we show these relations together with the associated inflationary observables for different choices of the parameter $\delta$. Their comparison to the Planck data confirm our initial choice of benchmark parameters and fixes the number of e-folds to be of order $N_{\star}\approx 53$ for $\delta\sim 10^{-11}$.

\section{Conclusion and Comments}\label{sec:conclusion}

We have considered a scalar portal scenario between the dark sector, composed of the inflaton and a pseudo scalar dark matter, and the visible sector, composed of a complex scalar field, assumed to be in thermal equilibrium with the standard model at high energies. Similarly to \cite{Boucenna:2014uma, King:2017nbl} the scalar potential of the dark sector encodes the inflationary potential and is invariant under a global $U(1)$ which is broken during inflation, whose pseudo-goldstone boson is assumed to be the dark constituent of our Universe. As setting the energy scale of inflation the inflaton-dark matter coupling to the inflation observables is constrained experimentally by astrophysical measurements. The production of dark matter is realized by annihilation of the visible pseudo-scalar into dark matter out of equilibrium through the inflaton portal term. We checked explicitly that the direct production of dark matter through inflaton decay at the reheating time doesn't dominate the dark matter production imposing further constraints on the parameter space. Finally we derived a detailed analysis of the number of e-fold as compared to experimental measurement and the reheating temperature obtained in our scenario confirming our choice of parameters. To put it in a nutshell, we obtain a scenario in which : $(i)$ the reheating temperature is of order $10^{11}$GeV, $(ii)$ the dark matter mass is lower than $\mathcal{O}(100)$ keV, $(iii)$ the inflaton portal necessary to get the measured relic density is of order $\delta\sim 10^{-(10-11)}$ and $(iv)$ a number of e-folds during inflation  $N_{\star}\sim 53$.

Assuming a quartic coupling in the visible sector of order $\lambda_{\sigma}\sim 0.1$, our scenario imposes the following hierarchy of scales
\begin{equation}
m_{\sigma}\gtrsim 10^{14}\mathrm{GeV}\gtrsim m_{\phi}\,,\quad \lambda_{\phi}\sim 10^{-13}\,,
\end{equation}
and thus
\be
v_{\phi}\sim 20 M_p\,,\qquad v_{\sigma}\gtrsim 10^{15}\mathrm{GeV}\,.
\ee

This letters thus opens the possibility that the inflaton provides the natural mediator between the dark and the visible sectors of our universe in a scenario consistent with astrophysical measurements and a minimal number of parameters without invoking any non minimal coupling to gravity.

We should make a few important remarks concerning the consistency of our dark matter production set up. First we assumed that the complex scalar $\Sigma$ was in close contact with the standard model, thus assuming that the pseudo-Goldstone $S_v$ is in thermal equilibrium with the standard model after the reheating and till rather low energies. We did not address in this paper how this could be the case. Such scalar could for instance be coupled to the neutrino sector as it is suggested in \cite{Boucenna:2014uma}, thus responsible for a high scale breaking of a Peccei-Quinn $U(1)$ symmetry. It could as well be responsible for the breaking of a flipped SU(5) or SO(10) down to SU(5) since its vev is required to be at the GUT scale, thus coupling directly to the standard model particles as fashioned for instance in \cite{Ahmad, Ellis:2014xda}. However, in any explicit model of interaction between the field $\Sigma$ and the standard model, one has to check that {\em(i)} the coupling of the pseudo goldstone to the visible sector is sufficiently strong to maintain it in equilibrium until low energies, {\em(ii)} being given such a strong coupling, the pseudo goldstone $S_v$ should escape any experimental detection (unless it is a prediction of the model) and {\em(iii)} the coupling of $S_v$ to the SM should not imply a too high coupling of the scalar $\sigma$ in a similar manner since it may significantly increase the reheating temperature.

We did not consider as well the possibility that the dark sector has a portal interaction to the scalar sector of the standard model since such term would interfere with the mass of the Higgs in the vacuum, due to the large vev of the inflaton. The coupling of such term would thus have to be fine-tuned to avoid such constraint, thus we ignored it in our analysis, although it could have interesting phenomenology at the time of the reheating, as pointed out in \cite{oleg}.

Finally, to be able to produce dark matter down to low energies, we made use of a visible pseudo scalar of mass of order $10$ MeV. Depending on the way such scalar couples to the standard model, it could have interesting experimental signatures.
\linebreak
\section{Acknowledgement}

I would like to thank Gaëtan Lafforgue-Marmet for collaboration at early stages of this work and interesting discussions as well as E. Dudas and T. Hambye for useful comments and suggestions. The work of L.H. is funded by the Belgian Federal Science Policy through the Interuniversity Attraction Pole P7/37.

\end{document}